\documentclass[%
 preprint, 
 amsmath,amssymb,
 aps, physrev,
letterpaper
]{revtex4-2}

\usepackage{dcolumn}
\usepackage{bm}
\usepackage{color}
\usepackage{graphicx}

\begin{document}

\preprint{APS/123-QED}

\title{Oscillatory liquid-metal flow in a channel under rapidly decaying applied magnetic field}

\author{Oleg Zikanov}
\email{Contact author: zikanov@umich.edu}
\affiliation{%
University of Michigan - Dearborn, 4901 Evergreen Rd., Dearborn, 48128, MI, USA
}%
\author{Haroon Ahmad}%
 \affiliation{%
 University of Michigan - Dearborn, 4901 Evergreen Rd., Dearborn, 48128, MI, USA
}%
\author{Sergey Smolentsev}%
 \affiliation{%
 Oak Ridge National Laboratory, Oak Ridge, 37932, TN, USA
}%

\date{\today}

\begin{abstract}
The channel flow of a liquid metal driven by a rapidly varying applied magnetic field is analyzed. The flow configuration, physical properties, and parameters correspond to a duct within a liquid-metal blanket of a nuclear fusion reactor under off-normal plasma conditions, such as plasma disruptions. The problem is solved numerically using a one-dimensional flow approximation. The longitudinal magnetic field, decaying at a typical rate on the order of 100 T/s, induces eddy currents that interact with a steady wall-normal magnetic field, generating the Lorentz force that drives the flow. Standing Alfvén waves are identified as the key mechanism controlling the liquid metal's response. These waves manifest as large-amplitude, gradually decaying oscillations of velocity, the induced magnetic field, and eddy currents. A parametric study predicts a severe response developing within the first few milliseconds of the event, with maximum flow velocities reaching several meters per second and Lorentz forces exceeding $10^9 \text{ N/m}^3$. Power-law approximations for the dependencies of the response characteristics on the flow parameters are developed. Finally, the effects of fluid compressibility and pressure waves are analyzed and found not to lead to a major modification of the flow evolution.
\end{abstract}

\maketitle


\section{Introduction}\label{sec:intro}

The breeder blanket surrounding the plasma chamber is a key element of a nuclear fusion reactor. It performs several critical functions: converting the energy of fusion-generated neutrons into heat, shielding external reactor components from neutron and electromagnetic loads, and breeding tritium fuel for the reaction. Many leading blanket concepts use liquid metals, typically a PbLi alloy \cite{smolentsev2024perspectives} or pure Li \cite{smolentsev6247477st}, as breeder and/or coolant materials.

Significant progress has been achieved in the understanding and computational modeling of liquid-metal blanket flows during normal, steady-state operation of the reactor (see, e.g., \cite{mistrangelo:2021b,mistrangelo:2021a,smolentsev2021physical} for recent reviews). In contrast, the processes occurring within the liquid metal in response to transient plasma events, such as disruptions, edge-localized modes, or vertical displacements \cite{hollmann2015status}, remain poorly understood. During such events, the state of the plasma and the electric currents flowing through it change very rapidly, on a typical time scale of $10^{-4}$ to $10^{-2}$~s. The resulting abrupt variation in the magnetic field within the blanket area induces strong eddy currents and Lorentz forces, resulting in a drastic change in the state of the system. While several studies have addressed electromagnetic forces, torques, and structural loads in solid blanket components during such events, the corresponding response of liquid components has received much less attention. 

The first systematic studies of liquid-metal blanket behavior during transient plasma events have recently been initiated within the ongoing SciDAC-5 program, "Center for Simulation of Plasma-Liquid Metal Interactions in Plasma-Facing Components and Breeding Blankets of a Fusion Power Reactor" \cite{SMOLENTSEV2026}. As part of this effort, the present study examines the flow generated by rapidly varying magnetic fields under conditions relevant to off-normal plasma transients. 

This work focuses on the plasma disruption event, which consists of a thermal quench followed by the termination of the toroidal electric currents that flow through the plasma during normal operation. This process is completed in about $10^{-2}$~s. The influence on the blanket is predominantly due to the decay of the poloidal magnetic field induced by plasma currents, dropping from an initial 0.5–1~T to practically zero. This occurs in the presence of a constant, strong toroidal magnetic field with a magnitude of 4–10~T. Preliminary estimates \cite{blanchard2019thermal,lei2022study,aduloju2026determination}, confirmed by recent studies \cite{Smolentsev2025,smolyanov2025exploratory,endeve2025full,SMOLENTSEV2026} and by the data of this work, predict a short-lived but extremely powerful response. This includes eddy currents with a density of $10^7$–$10^8$~A/m$^2$, associated Lorentz forces of $10^7$–$10^8$~N/m$^3$, and the acceleration of the liquid metal up to $10^3$–$10^4$~m/s$^2$. Consequently, the liquid metal experiences a uniquely rapid change of state, acquiring velocities of 1–10~m/s within a few milliseconds. The anticipated impact on reactor operation is not yet fully clear, but it may be significant. Possible consequences include transient pressure loads and electromagnetic forces on blanket structures, cyclic loading, and fatigue during repeated off-normal events, and modification of the electromagnetic environment coupled to the plasma. The extent of these effects remains to be quantified in integrated plasma–blanket simulations. 

We note that the analysis of the blanket response to transient plasma events must be conducted within the framework of a full resistive magnetohydrodynamics (MHD) model. As discussed in detail by \cite{smolyanov2025exploratory}, the quasi-static (inductionless) approximation commonly used in the analysis of liquid-metal blanket flows is inapplicable in the present case because the assumptions justifying this approximation are violated. The specific arguments supporting this statement are briefly outlined in Section \ref{sec:model} of this paper.

To date, the work on the response of the liquid-metal blanket to transient plasma has been primarily focused on plasma disruptions and on the flows within a poloidal duct—a central element in many currently pursued blanket designs, such as the DCLL (Dual Coolant Lithium Lead) concept developed in the U.S. \cite{smolentsev2024perspectives}. We briefly review these recent works before presenting the motivation for the present study.

A unidirectional, two-dimensional flow in a duct with walls of arbitrary electrical conductivity was analyzed in \cite{Smolentsev2025}. The flow was driven by a constant applied longitudinal pressure gradient and a longitudinal magnetic field decaying at a constant rate, in the presence of a constant transverse magnetic field. The study presented a family of analytical solutions for an asymptotic, fully developed flow, as well as numerical simulations of its transient behavior.  {The characteristic features of the response flow, such as antisymmetric velocity profiles and Hartmann-like boundary layers, were identified}.

The general applicability of conventional MHD models to describing liquid-metal blanket flows during transient plasma events was scrutinized in \cite{smolyanov2025exploratory}. The order-of-magnitude analysis demonstrated that many commonly used assumptions remain valid. In particular, the change of temperature due to heating by eddy currents is insignificant, allowing the physical properties of the liquid metal to be treated as constant. However, the effects of pressure waves on the flow evolution cannot be a priori excluded, prompting the authors to suggest further analysis. 

The numerical simulations in \cite{smolyanov2025exploratory} were conducted for a liquid metal in an elongated cuboid subjected to a constant transverse and an exponentially decaying longitudinal magnetic field.  {To isolate MHD interactions, the authors used a highly reduced hydrodynamic model that ignored viscous friction, non-linear advection, pressure gradients, and wall boundary conditions. Consequently, the recovered MHD flow effects were limited to the acceleration of the liquid metal by the Lorentz force and the velocity contribution in Ohm's law.}

More recently, \cite{endeve2025full} presented an extension of the Vertex-CFD code designed for full-induction MHD modeling of liquid-metal blanket flows. The ultimate objective of the work is the development of an integrated model that electromagnetically couples the unsteady plasma with the blanket flows.  {The model has been successfully verified against the analytical and numerical benchmark solutions of \cite{Smolentsev2025} and has demonstrated its ability to accurately reproduce the features of the response flow.}

Yet another study focusing on numerical methodologies \cite{Smolyanov2026} evaluated the performance of several modeling approaches, analyzing the adaptation of general-purpose CFD tools versus in-house codes for simulating full-MHD liquid-metal flows in a channel.

Despite the profound differences in problem setting and approach to analysis, the scenarios for the liquid-metal response emerging in the recent studies \cite{Smolentsev2025,smolyanov2025exploratory,endeve2025full} are qualitatively similar. A decaying poloidal field generates eddy currents primarily in the form of loops in the transverse plane. The interaction between these currents and the constant toroidal magnetic field generates antisymmetric Lorentz forces, which accelerate the fluid. The subsequent flow evolution shows several unique features. One of them is the quasi-one-dimensional structure: the velocity, Lorentz force, and induced magnetic field possess dominant components aligned with the longitudinal (poloidal) axis. The solution varies primarily along the direction of the imposed transverse (toroidal) magnetic field, whereas variations along the other two directions are substantially weaker. Furthermore, the distributions of the solution variables exhibit exceptionally thin Hartmann-like boundary layers near the walls perpendicular to the toroidal field \cite{Smolentsev2025,endeve2025full}. The eddy currents are concentrated, and the velocity and induced magnetic field have steep gradients within these layers.  {Finally, the reduced-hydrodynamics simulations of \cite{smolyanov2025exploratory} indicate a possible scenario where the accelerated liquid metal interacts with the wall-normal magnetic field to create reverse electric currents and Lorentz forces, initiating an oscillatory response.}

 {This work builds on the results of \cite{Smolentsev2025,smolyanov2025exploratory,endeve2025full}. Our goal is not a more elaborate and detailed numerical model, but a better understanding of the fundamental physical mechanisms that govern the flow behavior. Based on the findings that the response in a poloidal duct is an essentially one-dimensional process in which the walls normal to the toroidal magnetic field play a key role, we simplify the problem. Specifically, we employ a model of channel flow in which the solution is a function solely of time and the spatial coordinate perpendicular to the walls. As demonstrated below, this model is sufficient to generate results consistent with earlier findings while providing the much-needed insight into the underlying physics of the response.} 

 \section{Model}\label{sec:model}
 \subsection{Physical model}\label{sec:phys_model}
\begin{figure}
    \centering
    \includegraphics[width=0.5\textwidth]{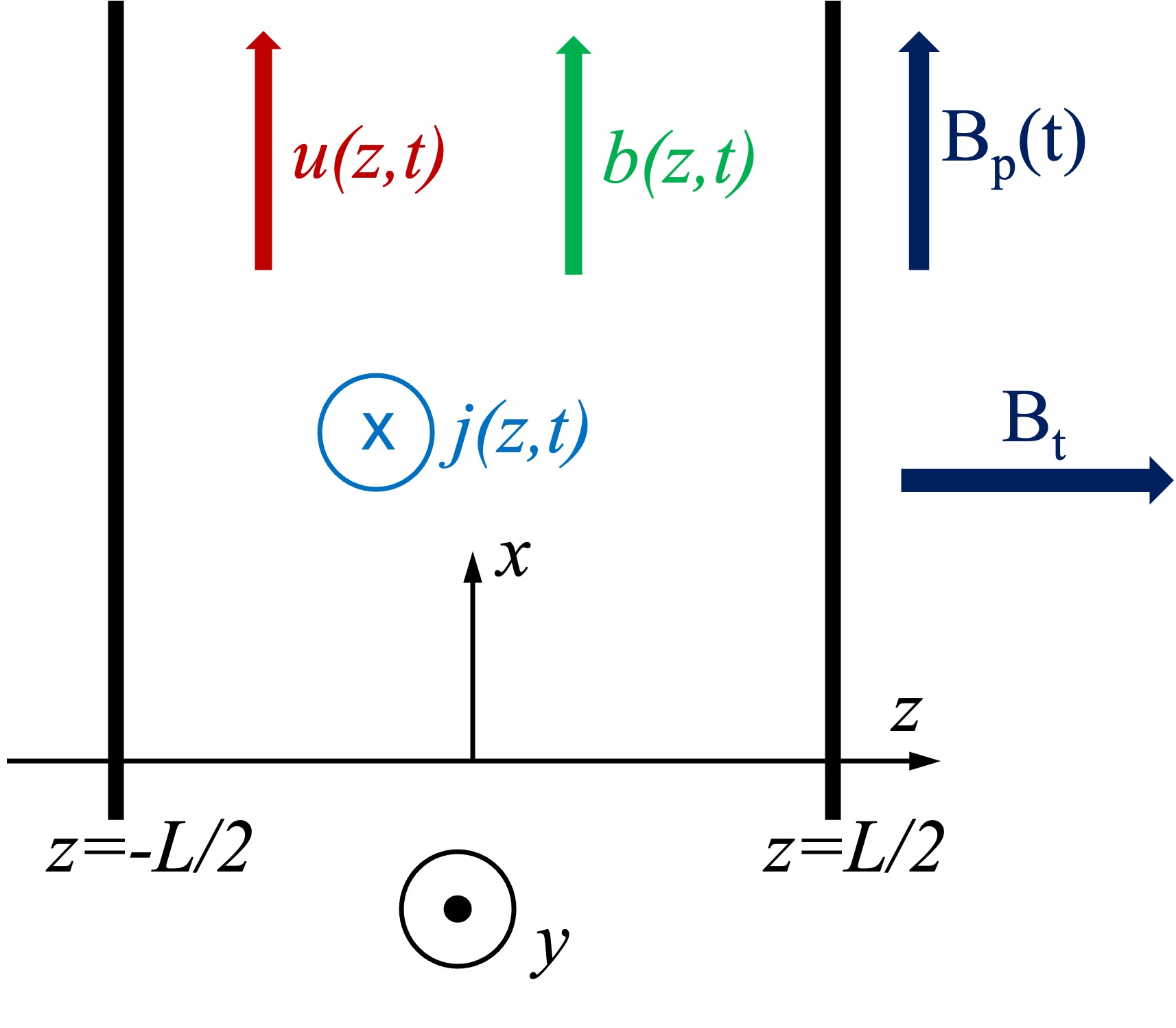}
    \caption{Schematic of the simplified flow geometry used in the study. The coordinates $x$, $y$, and $z$ correspond to the poloidal, radial, and toroidal directions of the fusion reactor coordinate system, respectively. A liquid metal is contained in a channel bounded by two electrically insulated walls at $z = \pm L/2$. The imposed magnetic field consists of a constant component $B_t\bm{e}_z$ and a decaying component $B_p(t)\bm{e}_x$. The orientations of the flow velocity $u$, induced magnetic field $b$, and induced eddy currents $j$ in the one-dimensional solution (see Section \ref{sec: problem} for an explanation) are indicated. Note that both the magnitude and direction (either as shown in the figure or opposite) of $u$, $b$, and $j$ vary with time and location within the channel.}\label{fig:geometry}
\end{figure}

 A flow of an electrically conducting, non-magnetizable fluid (a liquid metal) in a gap between two parallel, perfectly electrically insulating walls is analyzed (see Fig.~\ref{fig:geometry}). The fluid is initially at rest, with its subsequent motion caused by an imposed time-dependent magnetic field. The specific parameters of the problem are selected so that the solution represents, in an idealized manner, the evolution of the liquid-metal flow developing within a poloidal duct of the blanket of a magnetic-confinement nuclear fusion reactor during a typical plasma disruption event. 
 
 In particular, the wall-to-wall distance is $L\sim 0.1-0.2$ m, which corresponds to the typical width of poloidal ducts in many blanket concepts.  The imposed magnetic field 
 \begin{equation}
     \label{eq:B0} \bm{B}_0=B_t\bm{e}_z+B_p(t)\bm{e}_x
 \end{equation}
imitates the typical superposition of the time-independent toroidal field $B_t\sim 4-10$ T generated by the reactor's magnet coils and the time-dependent poloidal component $B_p$ generated by plasma current. During a plasma disruption event, $B_p$ decays from its normal operational value $B_p^0\sim 0.5-2$~T to nearly zero:
 \begin{equation}
     \label{eq:Bp} B_p(t)=B_p^0exp(-t/\tau),
 \end{equation}
where $\tau \sim 10^{-2}$ s is the typical timescale of the decay.  

The material properties of the two main candidates for the role of working fluid in a liquid metal blanket are selected for the analysis: PbLi alloy at a temperature of 300$^\circ$C and pure Li at 220$^\circ$C (see Table \ref{tab:properties}). 
\begin{table}[]
    \centering
    \begin{tabular}{c|l|l}
          Property& PbLi at 300$^\circ$C&Li at 220$^\circ$C\\\hline
          Density $\rho$, $kg/m^3$ & 9400&510\\ 
 Electric conductivity $\sigma$, $1/\Omega \cdot m$  & $0.79\cdot 10^6$ & $3.8\cdot 10^6$ \\ 
 Magnetic diffusivity $\eta=\sigma^{-1}\mu_0^{-1}$, $m^2/s$& 0.993&0.209\\
 Kinematic viscosity $\nu$, $m^2/s$ & $2.1\cdot 10^{-7}$ &$1.06\cdot 10^{-6}$ \\ 
Speed of sound $C$, $m/s$  & 1784 \cite{ueki2009acoustic} & 4512 \cite{blairs2007review}\\ \hline
\end{tabular}
    \caption{Fluid physical properties used in the study. If unspecified, the data are from \cite{zikanov2021}, where a compilation of physical properties of liquid metals from literature is reported. }
    \label{tab:properties}
\end{table}

One important conclusion of earlier studies \cite{Smolentsev2025, smolyanov2025exploratory, endeve2025full} is that the quasi-static (inductionless) approximation of the MHD model (see, e.g., \cite{davidson2016}), commonly used in the analysis of liquid-metal flows in reactor components \cite{mistrangelo:2021b,mistrangelo:2021a,smolentsev2021physical}, cannot be applied to the description of flows caused by transient plasma events. As explained in detail in \cite{smolyanov2025exploratory}, the reason for this is two-fold. First, with the typical velocity $U$ potentially being as high as several m/s, the magnetic Reynolds number $Rm\equiv UL/\eta$ can be $\sim 1$. Here, $L$ is the channel width, $\eta=(\sigma \mu_0)^{-1}$ is the magnetic diffusivity coefficient, $\sigma$ is the electric conductivity of the fluid, and $\mu_0$ is the magnetic permeability of vacuum. We note, however, that the argument based on a moderately large $Rm$ is not always sufficient. The practice of MHD flow simulations suggests that the quasi-static approximation remains acceptably accurate at $Rm\sim 0.1$ or even higher. 

The second, stronger, and more universally valid argument against the quasi-static model concerns the timescales involved \cite{smolyanov2025exploratory}. The typical magnetic diffusion time $\tau_m\sim L^2/\eta$ is comparable to the typical duration of the entire transient event. This invalidates the assumption of an instantaneous adjustment of the induced magnetic field and eddy currents to changes in the flow structure. 

 {Based on these arguments,} a full resistive MHD model that accounts for the complete two-way coupling between the motion of the electrically conducting fluid and the electromagnetic fields is used in the present study.

Other features of the model are based on the order-of-magnitude analysis of \cite{smolyanov2025exploratory}, which was conducted for the same typical values of $L$ and $\tau$ as shown above and the properties of PbLi listed in Table \ref{tab:properties}. It is easily verifiable, following the procedure of \cite{smolyanov2025exploratory}, that the conclusions are also valid for liquid Li.  In particular, it is shown in \cite{smolyanov2025exploratory} that the cumulative effect of the Joule heating by induced eddy currents is negligible, with a possible increase in temperature $\sim 1$ K or less. This means we can assume that the fluid temperature and the temperature-dependent physical properties, such as $\sigma$, $\eta$, $\nu$, or $C$ can be assumed constant without any significant loss of accuracy.  

The situation with density $\rho$ is more subtle. The high speed of sound $C$, and, thus, low isentropic compressibility coefficient $\beta_s\equiv \rho^{-1}(\partial \rho/\partial p)_s=\rho^{-1}C^{-2}$ of liquid metals implies that the pressure-induced fluctuations of density $\rho'$ are negligibly weak. The estimates of \cite{smolyanov2025exploratory} suggest the typical amplitude 
\begin{equation}
\label{eq:rhopert}|\rho'|<10^{-3}\rho.
\end{equation}
At the same time,  {it is suggested in \cite{smolyanov2025exploratory} that} a significant effect of compressibility cannot be excluded outright. The reason is that even at such a high $C$, the typical time of travel of pressure waves across the fluid domain is not much smaller than the typical time of the processes analyzed here. The assumption of infinitely fast propagation of pressure perturbations underlying the classical incompressibility model can be inaccurate. Since, as we will illustrate below, high-amplitude pressure variations develop in the system, this may affect the model's predictions. This question is explored further in the simulations presented below.

If the fluid is assumed to be incompressible, the governing equations are:
\begin{eqnarray}
\label{eq:incompr}    \nabla \cdot \bm{u} & = & 0\\
\label{eq:NSt}    \rho \left[ \frac{\partial \bm{u}}{\partial t} + (\bm{u}\cdot \nabla) \bm{u} \right] & = & -\nabla p + \nu \rho \nabla^{2}\bm{u} + \bm{J}\times (\bm{B}_0+\bm{b})\\
 \label{eq:induction-sep}  \frac{\partial \bm{b}}{\partial t} + (\bm{u}\cdot \nabla) \bm{b}  & = &  \left[ (\bm{B}_0+\bm{b})\cdot \nabla \right] \bm{u} + \eta \nabla^{2} \bm{b}-\frac{\partial \bm{B}_0}{\partial t},\\
  \label{eq:divb-sep}  \nabla \cdot\mathbf{b} &= & 0.\\
\label{eq:ampere}    \bm{J} & = & \frac{1}{\mu_{0}} \left( \nabla \times \bm{b} \right).
\end{eqnarray}
Here $\bm{u}$ and $p$ are the velocity and pressure fields, $\bm{J}$ is the density of the induced eddy currents, and $\bm{b}$ is the perturbation magnetic field induced internally during the event, such that the total magnetic field is $\bm{B}=\bm{B}_0+\bm{b}$.

When the incompressibility assumption is not used, we rely on the estimate of weak density perturbation \eqref{eq:rhopert} and negligible variation of temperature, and apply the linearized barotropic model
\begin{equation}
    \label{eq:barotropic} p'=\rho' C^2,
\end{equation}
where $p'$ and $\rho'$ are the perturbations with respect to the constants $p_0$, $\rho_0$ corresponding to the state of the fluid at the beginning of the event. The viscous attenuation of pressure waves is neglected because it occurs on the time scales of viscous dissipation, which are several orders of magnitude larger than the typical times of the flow evolution considered here. 

The hydrodynamic equations \eqref{eq:incompr}, \eqref{eq:NSt} are replaced by
\begin{eqnarray}
\label{eq:cont} \frac{\partial p}{\partial t} +\bm{u}\cdot \nabla p & = & -\rho_0C^2\nabla\cdot \bm{u} ,\\
\label{eq:NStcomp}    \rho_0 \left[ \frac{\partial \bm{u}}{\partial t} + (\bm{u}\cdot \nabla) \bm{u} \right] & = & -\nabla p + \nu \rho_0 \nabla^{2}\bm{u} + \bm{J}\times (\bm{B}_0+\bm{b}),
\end{eqnarray}
where we substitute the full pressure field $p$  for $p'$ under the derivatives, and $\bm{u}$ includes the velocity perturbations $\bm{u'}$ due to the pressure waves.

\subsection{Non-dimensional parameters}\label{sec:nondim_param}
The non-dimensional parameters of the problem are listed here for illustrative purposes. The problem is solved in dimensional units. The main reason is that the phenomenon considered in this paper occurs within a limited range of variable parameters ($L$, $\tau$, $B_t$, $B_p^0$) and for just two specific materials (PbLi and Li). Observing the effects of the dimensional parameters directly, rather than through non-dimensional combinations, creates a clearer and practically more useful picture of the behavior. 

Our version of non-dimensionalization is based on the Alfv\'{e}n velocity 
\begin{equation}
\label{eq:alf_vel}    U_A\equiv B_t\mu_0^{-1/2}\rho^{-1/2}
\end{equation}
as the typical velocity scale and  the decay time of the poloidal field $\tau$ as the time scale. The channel width $L$ is used as the typical length scale. This selection is appropriate for the present analysis because, as shown below, the dominant oscillatory response is associated with standing Alfvén waves propagating across the channel between the two walls. At the same time, the Hartmann-layer thickness  {$\delta_{Ha}=L/Ha$, where $Ha$ is the Hartmann number defined below,} remains an important local scale controlling near-wall gradients, current concentration, and dissipation. The magnetic field and hydrodynamic pressure are scaled by $B_t$ and $B_t^2/\mu_0$, respectively. 

The resulting non-dimensional parameters are the hydrodynamic and magnetic Reynolds numbers, the Strouhal number, and the ratio between the poloidal and toroidal magnetic fields:
\begin{equation}
    \label{eq:nondim_par} Re\equiv \frac{B_t L}{\nu (\mu_0 \rho)^{1/2}}, \: Rm\equiv \frac{B_tL\sigma \mu_0^{1/2}}{\rho^{1/2}}, \: Sr\equiv \frac{L\mu_0^{1/2}\rho^{1/2}}{B_t \tau}, \: \Gamma\equiv \frac{B_p^0}{B_t}.
\end{equation}
In the following discussion, we will also use the Hartmann number, defined in our units as 
\begin{equation} 
\label{eq:Ha}Ha\equiv  LB_t (\sigma/\rho \nu)^{1/2}=(ReRm)^{1/2}.
\end{equation}
The Hartmann number is commonly used in liquid-metal MHD, with $Ha^2$ representing the typical ratio between the Lorentz and viscous friction forces. 

If the compressible flow model is used, one must add the Mach number 
\begin{equation}
    \label{eq:Mach} M\equiv U_AC^{-1}= B_t\mu_0^{-1/2}\rho^{-1/2} C^{-1} .
\end{equation}

\subsection{Problem statement}\label{sec: problem}
A poloidal duct of a liquid-metal blanket is an elongated cuboid with cross-sectional dimensions of $\sim 0.1-0.2$~m and a length of $\sim 1-10$~m. As we have already mentioned in the Introduction, the previous attempts at analyzing the response in such geometries \cite{Smolentsev2025,smolyanov2025exploratory,endeve2025full} have shown qualitatively similar system behaviors. The key feature is the development of flow structures near the walls normal to the constant toroidal component of the imposed magnetic field. These structures include peaks of longitudinal velocity and, as clearly identified in \cite{Smolentsev2025}, thin boundary layers characterized by sharp gradients of velocity and induced magnetic field, and concentration of the induced eddy currents. The spatial pattern of the flow is dominated by variations along the wall-normal coordinate, whereas variations along the other two axes are significantly weaker.

Based on these observations, we apply the approximation in which the fluid is contained in a channel between two parallel, perfectly insulating electrical walls (see Fig.~\ref{fig:geometry}). The secondary effects, in particular those of the side and end walls, and of the finite electrical conductivity of the walls, are left to future studies. The solution is represented as one-dimensional, depending only on the wall-normal coordinate $z$ and time $t$. As demonstrated below, the advantages of this approach include computational accuracy, simplicity, low computational cost, and, most importantly, a clear insight into the primary physical mechanisms of the flow, free from secondary effects.

For the main part of the analysis, the fluid is assumed to be incompressible. We assume the solution in the form $\bm{u}=u(z,t)\bm{e}_x$, $\bm{b}=b(z,t)\bm{e}_x$, $\bm{J}=j(z,t)\bm{e}_y$, $p=p(z,t)$. The system \eqref{eq:incompr}-\eqref{eq:ampere} simplifies to
\begin{eqnarray}
\label{eq:1da_inc}\frac{\partial u}{\partial t} & = & \nu \frac{\partial^2u}{\partial z^2}+\frac{B_t}{\mu_0 \rho}\frac{\partial b}{\partial z},\\
\label{eq:1db_inc}\frac{\partial b}{\partial t} & = & B_t \frac{\partial u}{\partial z}+\eta \frac{\partial^2 b}{\partial z^2}+\frac{B_p^0}{\tau}\exp\left(-\frac{t}{\tau}\right).
\end{eqnarray}
Ampere's law \eqref{eq:ampere} becomes
\begin{equation}
\label{eq:1dcurr_inc}   j=\frac{1}{\mu_0}\frac{\partial b}{\partial z}.  
\end{equation}
The pressure distribution is determined by the $z$-component of the momentum equation \eqref{eq:NSt}, which is reduced to the balance between the pressure gradient and the Lorentz force:
\begin{equation}
    \label{eq:pressure_inc} \frac{\partial p}{\partial z}=-j\left(B_p(t)+b\right)=-\frac{1}{\mu_0}\frac{\partial b}{\partial z}\left(B_p(t)+b\right).
\end{equation}
The boundary conditions are:
\begin{eqnarray}
\label{eq:bc_u_inc} u=0 & \textrm{ at } & z=\pm L/2,\\
\label{eq:bc_b_inc} b=0 & \textrm{ at } & z=\pm L/2.
\end{eqnarray}
We note here an additional benefit of the one-dimensional problem statement. It leads to an exact boundary condition \eqref{eq:bc_b_inc} for the induced magnetic field imposed at the non-conducting walls.  {In the general case, no such condition is available, and the solution for the induced magnetic fields  must be extended outside the flow domain \cite{smolyanov2025exploratory}.} 

Further conditions include the automatically satisfied condition of zero net current in the $y$-direction (necessary if electrically insulating walls crossing the $y$-axis are presumed) $ \int_{-L/2}^{L/2} j dz =0$, and the reference value for pressure, which we take as $p_{z=-L/2}=0$.

The volume-averaged energy balance used for analysis can be obtained by multiplying \eqref{eq:1da_inc} by $u$, \eqref{eq:1db_inc} by $b$ and integrating in $z$:
\begin{eqnarray}
\label{eq:bal_a_inc} \frac{d \textrm{KE}}{dt} & = & -E+N,\\ 
\label{eq:bal_b_inc} \frac{d \textrm{ME}}{dt} & = & -E_m-N+S,
\end{eqnarray}
where 
\begin{equation}
  \label{eq:en_def_inc}  \textrm{KE}=\frac{\rho}{2L}\int_{-L/2}^{L/2} u^2 dz, \: \: \: \textrm{ME}=\frac{1}{2\mu_0L}\int_{-L/2}^{L/2} b^2 dz
\end{equation}
are the volume-averaged kinetic and magnetic energies, 
\begin{equation}
 \label{eq:terms1_inc}     E=\frac{\rho\nu}{L}\int_{-L/2}^{L/2} \left(\frac{\partial u}{\partial z}\right)^2 dz, \: \: \: E_m=\frac{1}{\sigma \mu_0^2 L}\int_{-L/2}^{L/2} \left(\frac{\partial b}{\partial z}\right)^2 dz
\end{equation}
are the rates of viscous and Joule dissipation, 
\begin{equation}
 \label{eq:terms2_inc}   N=\frac{B_t}{L\mu_0}\int_{-L/2}^{L/2} u\frac{\partial b}{\partial z}dz
\end{equation}
is the rate of energy transfer between the magnetic field and the flow, and 
\begin{equation}
 \label{eq:terms3_inc}     S=\frac{B_p^0\exp(-t/\tau)}{L\mu_0 \tau}\int_{-L/2}^{L/2} b dz
\end{equation}
is the source term representing the energy input from the time-varying imposed magnetic field.

 {The configuration of the problem, namely a flow of an electrically conducting fluid crossing the magnetic field lines $B_t\bm{e}_z$, invites analysis in terms of  Alfvén waves. This is a classical MHD phenomenon (see, e.g., \cite{davidson2016}), more commonly considered in the astrophysical context. The waves are caused by magnetic tension arising in response to deformation of the magnetic field lines by a moving, electrically conducting fluid.}

 {We will consider the actual realization of such waves in our system in Section~\ref{sec:typical_solutions}. For now, we observe that, at asymptotically high hydrodynamic and magnetic Reynolds numbers, the non-diffusive version of \eqref{eq:1da_inc} and \eqref{eq:1db_inc} leads to the hyperbolic equations:}
\begin{eqnarray}
\label{eq:alf_a}\frac{\partial^2 u}{\partial t^2} & = & U_A^2\frac{\partial^2 u}{\partial z^2},\\
\label{eq:alf_b}\frac{\partial^2 b}{\partial t^2} & = & U_A^2 \frac{\partial^2 b}{\partial z^2}-\frac{B_p^0}{\tau^2}\exp\left(-\frac{t}{\tau}\right),
\end{eqnarray}
with the Alfvén velocity \eqref{eq:alf_vel} as the wave speed. The presence of channel walls perpendicular to the direction of wave propagation implies the realization in the form of standing waves, where the typical time period of oscillations corresponds to the wave traveling from one wall to the other and back, $T_A=2L/U_A$. 

\subsection{Numerical method}\label{sec:method}
The problem \eqref{eq:1da_inc}--\eqref{eq:bc_b_inc} does not have a simple analytical solution in the non-stationary case. Although it could, in principle, be solved using the method of separation of variables, the presence of extremely thin boundary layers in the solution (as demonstrated below) poses a significant challenge. Specifically, this approach would involve a slowly converging expansion series and would suffer from Gibbs oscillations. Therefore, a more straightforward and reliable approach is to solve the problem numerically.

The numerical solution is obtained using the PDE solver \emph{pdepe} of MATLAB \cite{mathworks_pdepe}. The method of lines is used. The second-order Skeel-Berzins spatial discretization \cite{skeel1990method} is applied. The resulting system of ordinary differential equations is solved by the MATLAB solver for stiff systems \emph{ode15} \cite{ShampineReichelt1997}, which uses implicit time integration and adapts the scheme order and integration time step to achieve the target absolute and relative tolerances.

The solution was verified in direct comparison with the solution obtained using the second-order finite-difference scheme developed in \cite{Krasnov2011,Krasnov2023} and adapted to flows with the full MHD effect in \cite{Smolyanov2026}. A sensitivity study was conducted to determine the spatial grid parameters (size and non-uniformity) and the tolerances of time discretization on the solution, which ensure solution accuracy. The results presented below are obtained using a spatial grid of 500 points and near-wall clustering
\begin{equation}
   \label{eq:tanh} z =\frac{L}{2} \frac{\tanh{(A_z\xi)} } {\tanh{(A_z)}},
\end{equation}
where $-1\le \xi\le 1$ is the transformed coordinate in which the grid is uniform, and $A_z=5.0$ is the clustering parameter. As an example, in the representative solution described in Section \ref{sec:typical_PbLi},  the smallest (nearest to the wall) grid cell has a size of $1.9\times 10^{-7}$ m.  The absolute and relative tolerances of time discretization are set  to $10^{-7}$ and $10^{-5}$, respectively.

\section{Representative solutions} \label{sec:typical_solutions}
A detailed discussion of representative solutions is provided in this section. The same system parameters as in \cite{smolyanov2025exploratory} are selected: $L=0.1$ m, $\tau=0.006$ s, $B_t=10$ T, $B_p^0=0.5$ T. While the flow evolution during the first 0.05 s of the response is simulated, only the first 0.01 s is illustrated and discussed below. The subsequent evolution of the flow is predominantly characterized by gradual decay. 

\subsection{PbLi flow}\label{sec:typical_PbLi}
\begin{figure}
    \centering
    \includegraphics[width=0.8\textwidth]{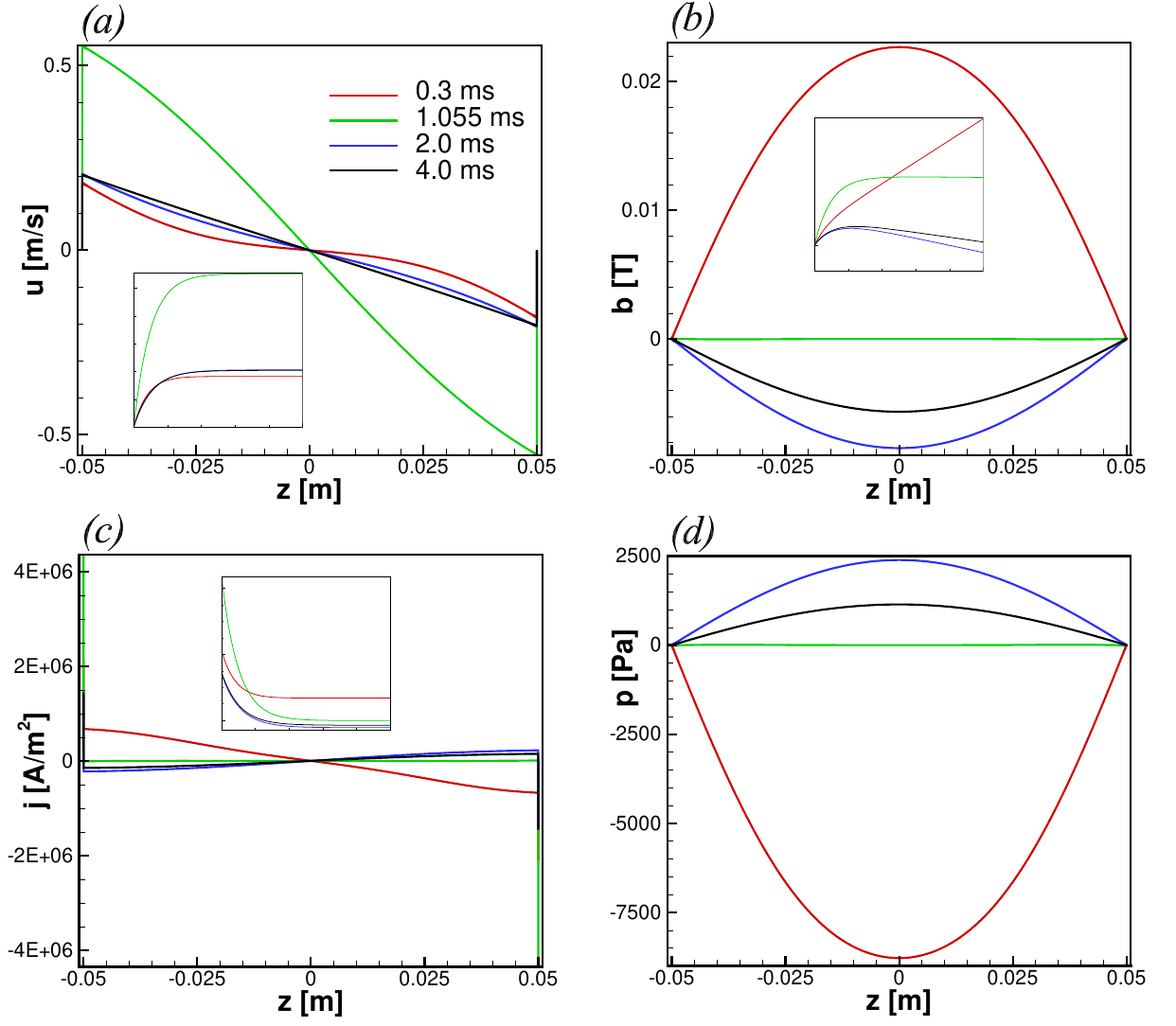}
    \caption{Solution for PbLi at $L=0.1$ m, $\tau=0.006$ s, $B_t=10$ T, $B_p^0=0.5$ T.  Snapshots of profiles of velocity $u$, induced magnetic field $b$, induced eddy current $j$ , and pressure $p$ at time moments $t=3\times 10^{-4}$ s, $1.055\times 10^{-3}$ s, $2\times 10^{-3}$ s, and $4\times 10^{-3}$~s are shown. The inserts in the plots for $u$, $b$, and $j$ show the distributions within a layer of width $5\times 10^{-5}$ m near the wall $z=-L/2$. Similar (symmetric for $b$ and antisymmetric for $u$ and $j$) distributions appear at the wall $z=L/2$. Plot \emph{(c)} also illustrates the distribution of the streamwise Lorentz force component $f_x=jB_t$.}
    \label{fig:rep_PbLi_profiles}
\end{figure}

\begin{figure}
    \centering
    \includegraphics[width=0.8\textwidth]{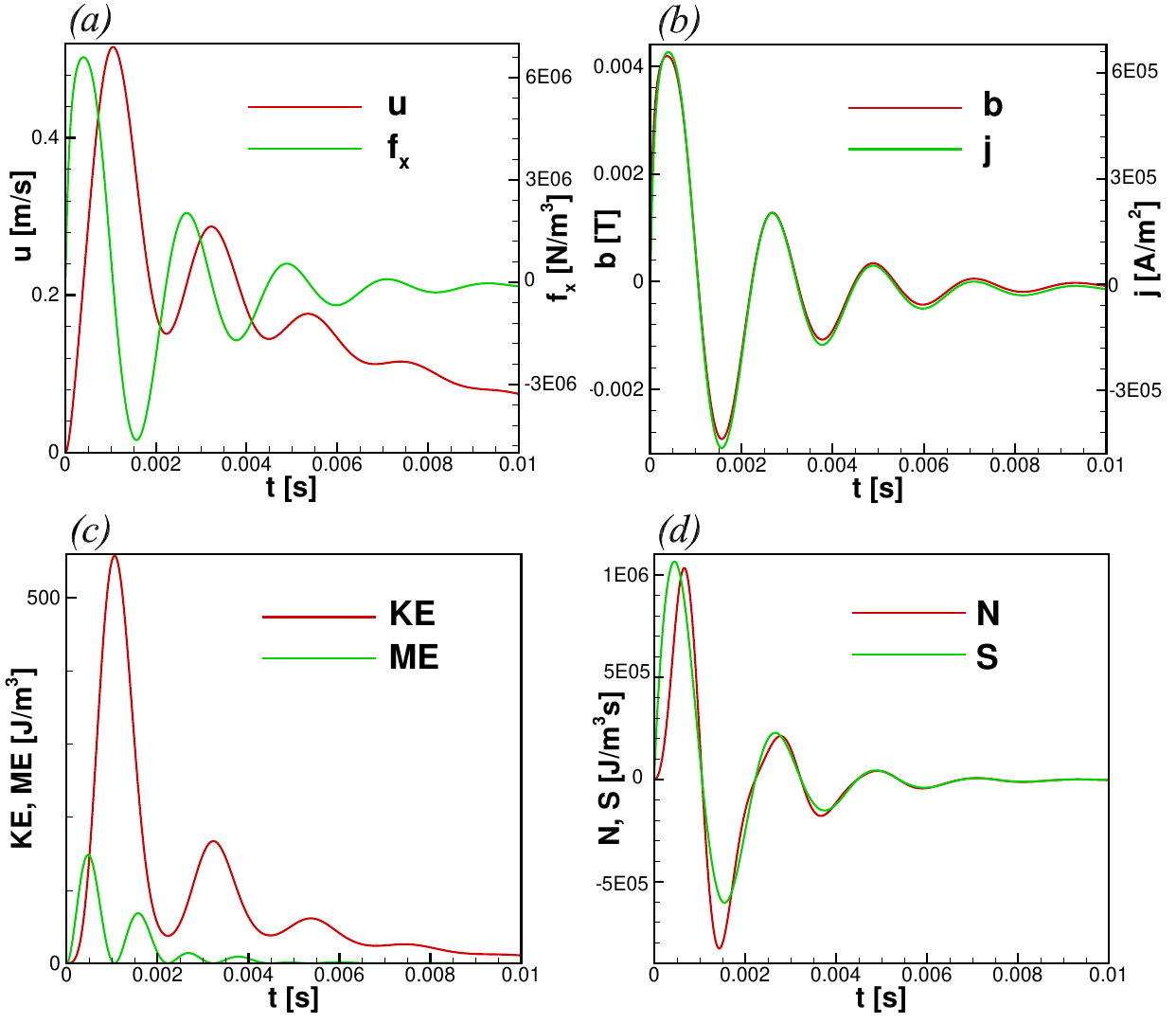}
    \caption{Solution for PbLi at $L=0.1$ m, $\tau=0.006$~s, $B_t=10$~T, and $B_p^0=0.5$~T.  Time evolution of velocity $u$, Lorentz force $f_x=jB_t$, induced magnetic field $b$, and electric current $j$ at the point $z=-0.45L$ is shown in \emph{(a)} and \emph{(b)}. Terms of the energy balance equations \eqref{eq:bal_a_inc}, \eqref{eq:bal_b_inc} are shown in \emph{(c)} and \emph{(d)}. }
    \label{fig:rep_PbLi_evolution}
\end{figure}

For the PbLi flow, the Alfv\'{e}n velocity is $U_A=92$~m/s, and the values of the non-dimensional parameters are $Re=4.38\times 10^7$, $Rm=9.13$, $Sr=0.181$, $\Gamma=0.05$, and $Ha=2.00\times 10^4$. The results obtained for this flow are illustrated in Figs.~\ref{fig:rep_PbLi_profiles} and \ref{fig:rep_PbLi_evolution}. 

The profiles of solution variables shown in Fig.~\ref{fig:rep_PbLi_profiles} exhibit features qualitatively similar to those observed in the earlier studies \cite{Smolentsev2025,smolyanov2025exploratory,endeve2025full}. In particular, these include the symmetry of the solution: the distributions of $u$, $j$, and the Lorentz force $f_x=jB_t$ are antisymmetric with respect to $z=0$, while the distributions of $b$ and $p$ are symmetric. The similarity of the $b$ and $p$ profiles, up to reflection symmetry, is consistent with \eqref{eq:pressure_inc}. Since $|b|\ll |B_p|$ in our solution, it follows that $p\approx -B_p(t)\mu_0^{-1}b$.

The flow development is very fast. After just $3\times 10^{-4}$~s, the electric current density exceeds $4\times 10^6$~A/m$^2$, while the velocity grows above 0.2~m/s. The subsequent evolution of the flow is characterized by a rapid and non-monotonic change in all variables, a phenomenon that will be addressed shortly.

A striking feature of the velocity and electric current profiles is the presence of extremely thin (with a thickness of about $ 2 \times 10^{-5}$~m) boundary layers, where the velocity adjusts to the no-slip condition, the induced magnetic field has sharp gradients, and the strongest $y$-directed electric current $j$ flows.  {Similar layers were identified in the steady-state asymptotic solution of \cite{Smolentsev2025} and numerical solutions of \cite{Smolentsev2025,endeve2025full,Smolyanov2026}.} Akin to the classical Hartmann flow, the boundary layer thickness is proportional to $\delta=L/Ha$, specifically on the order of $5\delta$ (see \cite{Smolyanov2026} for the explicit analytical solution and illustrations for the steady-state channel flow). Given that $\delta=5\times 10^{-6}$~m in our representative example, the results illustrated in Fig.~\ref{fig:rep_PbLi_profiles} align well with the findings of \cite{Smolentsev2025,endeve2025full,Smolyanov2026}. 

 {This scaling of the boundary layer thickness may appear improbable at first, as the classical Hartmann boundary layer is an equilibrium structure formed by a balance between Lorentz forces and viscous friction. While one expects such a balance in the asymptotic steady-state solutions of \cite{Smolentsev2025}, it may seem unexpected} in a highly transient case like ours. In fact, the typical viscous diffusion time across the boundary layer is only $(5\delta)^2/\nu \approx 3\times 10^{-3}$~s, meaning there is no contradiction.

We stress that the computational grid used in our study provides sufficient resolution of the boundary layers. Specifically, 19 mesh points are located within the near-wall layer of thickness $\delta$, while the entire near-wall interval of width $5\delta$ is resolved by 47 points.

The evolution of the system is illustrated in Fig.~\ref{fig:rep_PbLi_evolution}. The general scenario is qualitatively similar to that reported in \cite{smolyanov2025exploratory}, although the picture presented by our solution is much more accurate because it relies on the complete hydrodynamic model and a fine-resolution numerical approximation. The defining characteristic of this process is the strong and rapid fluctuations of all variables, including the point signals of $u$, $f_x$, $b$, and $j$ (see Figs.~\ref{fig:rep_PbLi_evolution}a, b) and the integrated terms of the energy balance equation (see Figs.~\ref{fig:rep_PbLi_evolution}c, d).

A detailed scenario explaining the fluctuations is as follows  {(a qualitatively similar explanation was proposed based on a reduced hydrodynamics model in \cite{smolyanov2025exploratory}).} The initial stage of flow development is characterized by acceleration caused by the rapid decay of $B_p(t)$ and the associated growth of the induced magnetic field $b>0$ (per Lenz's law). The induced current $j$ peaks near the walls and is directed in the positive $y$-direction for $z<0$ and the negative $y$-direction for $z>0$. The resulting Lorentz force $f_x=B_tj$ accelerates the liquid metal in the positive $x$-direction at $z<0$ and in the negative $x$-direction at $z>0$. This stage is manifested by the growth of the point values of $u$, $b$, $f_x$, and $j$ (see Figs.~\ref{fig:rep_PbLi_evolution}a, b) as well as the kinetic and magnetic energies (see Fig.~\ref{fig:rep_PbLi_evolution}c). The positive, growing values of the integrated energy input $S$ and energy transfer $N$ (see \eqref{eq:terms2_inc}, \eqref{eq:terms3_inc}) demonstrate the main direction of the energy flow during this stage: from the imposed poloidal magnetic field to the induced field, and subsequently to the  {accelerating flow of the liquid metal.} The state of the solution during this stage is illustrated by the profiles at $t=0.3\times 10^{-3}$~s in Fig.~\ref{fig:rep_PbLi_profiles}.

The growing flow velocity leads to an effect that curtails this growth and brings the initial stage to an end. The underlying mechanism involves the electrically conducting fluid moving across the toroidal field $B_t\bm{e}_z$ and inducing electric currents that oppose those created via the Lenz effect by the decaying poloidal field. This can be seen from the $\bm{u}\times \bm{B}$ term in Ohm's law. Alternatively, one can differentiate \eqref{eq:1db_inc} with respect to $z$ to obtain, from the first two terms:
\begin{equation}
\label{eq:explain} \mu_0\frac{\partial j}{\partial t}= B_t\frac{\partial^2 u}{\partial z^2}+\cdots
\end{equation}
The sign of $\partial^2 u/\partial z^2$ is such that it obviously counteracts $j$ (see the curves for $u$ and $j$ at $t=0.3\times 10^{-3}$~s in Fig.~\ref{fig:rep_PbLi_profiles}). 

The growth of the opposite electric currents occurs against the background of a reducing induction driven by $d B_p/dt$. As a result, $j$ decreases in amplitude and subsequently changes sign (see the point signals in Fig.~\ref{fig:rep_PbLi_evolution}b and the evolution of the $j$ and $b$ profiles in Figs.~\ref{fig:rep_PbLi_profiles}b, c). The Lorentz force $f_x$ consequently decreases in magnitude and changes its sign, so it begins to decelerate the flow (see Fig.~\ref{fig:rep_PbLi_evolution}a). The magnetic energy $\textrm{ME}$, followed by the kinetic energy $\textrm{KE}$, starts to decrease (see Fig.~\ref{fig:rep_PbLi_evolution}c). The reduction and eventual sign change of $b$ lead to a corresponding decrease and sign change in both the energy source term $S$ and the energy transfer term $N$.

As the flow decelerates, the effect of the velocity on the induced electric current decreases. Consequently, the growth of $b$ and $j$ generated by the variation of $B_p(t)$ reappears approximately after $t=10^{-3}$~s (see Fig.~\ref{fig:rep_PbLi_evolution}), albeit at a lower rate due to the reduced amplitude of $dB_p/dt$. The subsequent evolution of the flow can be described as a sequence of gradually diminishing, alternating stages of growth and reversal.

The analysis of the solution data shows that both the peaks and the local minima of the kinetic energy $\textrm{KE}(t)$ occur at the same moments as the local minima (with nearly zero values) of the magnetic energy $\textrm{ME}(t)$ (see Fig.~\ref{fig:rep_PbLi_evolution}). The energy transfer rate $N(t)$ and the energy source $S(t)$ both cross zero at these same instances. For example, this is observed at the first peak of $\textrm{KE}$ at $t = 1.055 \times 10^{-3}$ s and at the first minimum of $\textrm{KE}$ at $t = 2.227 \times 10^{-3}$ s. The profiles of the solution variables demonstrate that these moments, which separate the stages of growth from those of decay, correspond to a nearly zero induced magnetic field across the entire channel (see the profile at $t = 1.055 \times 10^{-3}$ s in Fig.~\ref{fig:rep_PbLi_profiles}b as an example).

 {There is also a shorter version of the explanation of the observed flow evolution. This interpretation is suggested for the first time in this paper. The problem can be viewed in terms of Alfvén waves in the $z$-direction; the fluctuations of the solution variables are then a manifestation of these waves. The entire scenario described above corresponds to the classical Alfvén wave scenario. A flow of liquid metal crossing the lines of the toroidal magnetic field $B_t\bm{e}_z$ distorts these lines by adding the field $b\bm{e}_x$ corresponding to the opposite electric currents induced by the flow. This results in a restoring magnetic tension force and leads to oscillations. The unique features observed in our system are the presence of electrically insulating walls, which causes the formation of standing waves, and the continuous forcing of the process by the decaying poloidal field $B_p(t)\bm{e}_x$.}

 {To verify this interpretation, we estimate the period of oscillations, which can be measured as the typical time interval $T_n$ between the peaks of the solution variables in Figs.~\ref{fig:rep_PbLi_evolution}a,b. The result is $T_n\approx 2.2\times 10^{-3}$~s initially, slowly decreasing over time due to the decay shift. This value is in agreement with the theoretical period of the Alfvén wave, $T_a = 2L/U_A = 2.17\times 10^{-3}$~s. The analogy is, of course, incomplete, since our solution also exhibits the effects of viscosity and finite magnetic diffusivity. The latter is especially significant due to the moderate magnetic Reynolds number, $Rm = 9.13$.}

\subsection{Li flow}\label{sec:typical_Li}
When the fluid is pure Li, the representative case corresponds to the Alfvén velocity $U_A=395$~m/s and the non-dimensional parameters $Re=3.73\times 10^7$, $Rm=189$, $Sr=0.0422$, $\Gamma=0.05$, and $Ha=8.38\times 10^4$. 

\begin{figure}
    \centering
    \includegraphics[width=0.8\textwidth]{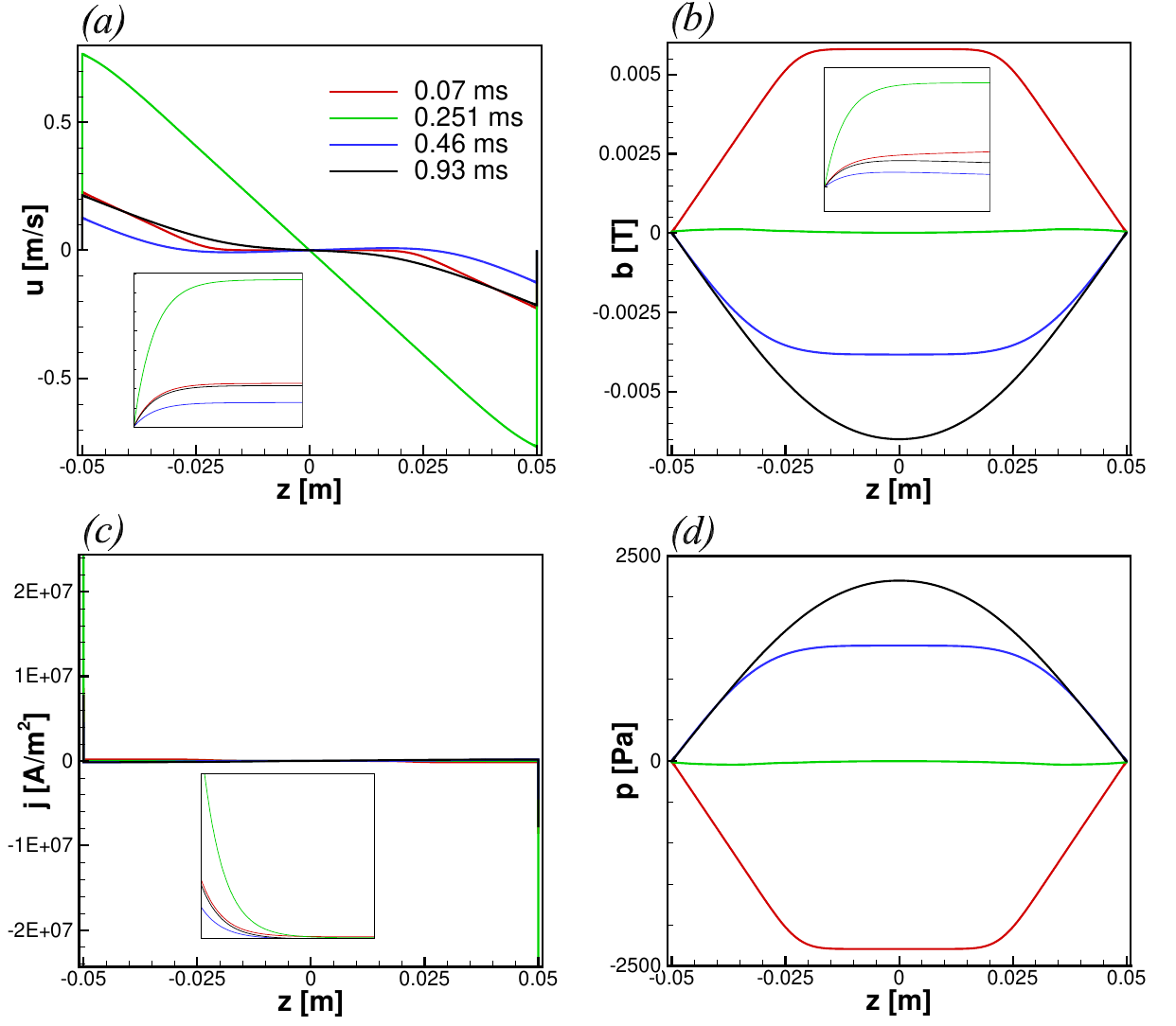}
    \caption{Solution for Li at $L=0.1$ m, $\tau=0.006$ s, $B_t=10$ T, $B_p^0=0.5$ T.  Snapshots of profiles of velocity $u$, induced magnetic field $b$, induced eddy current $j$ , and pressure $p$ at time moments $t=7\times 10^{-5}$ s, $2.51\times 10^{-4}$ s, $4.6\times 10^{-4}$ s, and $9.3\times 10^{-4}$~s are shown. The inserts in the plots for $u$, $b$, and $j$ show the distributions within a layer of width $ 10^{-5}$ m near the wall $z=-L/2$. Similar (symmetric for $b$ and antisymmetric for $u$ and $j$) distributions appear at the wall $z=L/2$. Plot \emph{(c)} also illustrates the distribution of the streamwise Lorentz force component $f_x=jB_t$.}
    \label{fig:rep_Li_profiles}
\end{figure}

The distributions of the solution variables across the channel are shown in Fig.~\ref{fig:rep_Li_profiles}. We note that the time instances for the profiles are selected to correspond to approximately the same stages of flow evolution as those shown in Fig.~\ref{fig:rep_PbLi_profiles} for the PbLi case. Furthermore, the close-ups near the wall at $z=-L/2$ show a thinner layer of $10^{-5}$~m to account for the larger $Ha$ in the Li flow case.

We see a clear qualitative similarity between the Li and PbLi cases. This similarity includes the symmetry and the presence of boundary layers of thickness $\delta\approx 5L/Ha$, within which the velocity adjusts to the no-slip condition, the induced magnetic field has sharp gradients, and the strongest spanwise current $j$ flows. The main quantitative difference is that the maximum velocity $u$ is slightly larger, while the maximum magnitude of the induced magnetic field is 3–4 times smaller than those in the PbLi flow. This difference can be attributed primarily to the much lower density of Li. Because the fluid acceleration produced by the Lorentz force is stronger, the velocity reaches a magnitude, at which the motion of the liquid metal across the magnetic field lines begins to significantly reduce the current and curtail further growth, faster and at a smaller amplitude of $b$.

\begin{figure}
    \centering
    \includegraphics[width=0.8\textwidth]{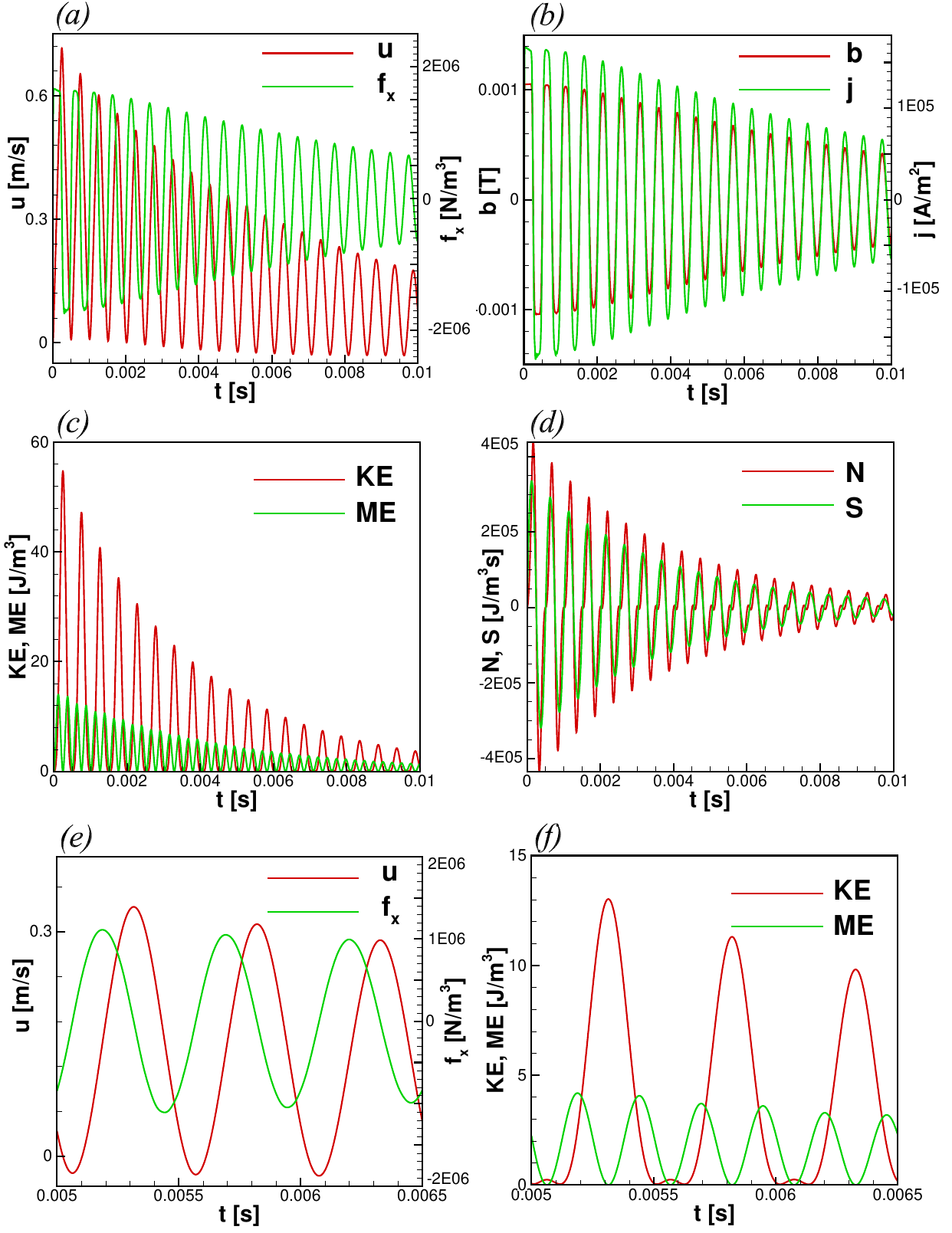}
    \caption{Solution for Li at $L=0.1$ m, $\tau=0.006$~s, $B_t=10$~T, and $B_p^0=0.5$~T. The time evolution of velocity $u$, Lorentz force $f_x=jB_t$, induced magnetic field $b$, and electric current $j$ at the point $z=-0.45L$ is shown in (a) and (b). Terms of the energy balance equations \eqref{eq:bal_a_inc} and \eqref{eq:bal_b_inc} are shown in (c) and (d). Panels (e) and (f) show the curves from (a) and (b) zoomed in over the time interval between 5 and 6.5 ms.}
    \label{fig:rep_Li_evolution}
\end{figure}

The flow evolution illustrated in Fig.~\ref{fig:rep_Li_evolution} is also qualitatively similar to that of the PbLi flow. We see alternating stages of growth and decay of solution variables. As before, the peaks and local minima of $\textrm{KE}$ occur simultaneously with minima of $\textrm{ME}$ and zeros of $N$ and $S$. These events correspond to $b\approx 0$ in the entire channel (see the profile at $t=0.251\times 10^{-3}$~s, which is the first peak of $\textrm{KE}$ in Fig.~\ref{fig:rep_Li_profiles}b).  We conclude that the scenario of flow evolution described above remains valid. 

Moreover, the Alfvén wave interpretation of the solution is much more direct and accurate in the Li flow case. The solution variables exhibit practically regular oscillations with a period $T_n$ in nearly perfect agreement with the theoretical prediction $T_a = 2L/U_A = 5.06\times 10^{-4}$~s. This greater validity can be explained by the much higher value of $Rm$ in the Li flow case.

\section{Effect of system parameters} \label{sec:effect_parameters}
Multiple simulations were carried out at various reactor-relevant combinations of the parameters. All of them showed realizations of the general scenario described above. The focus of the parametric study is, therefore, on the quantitative characteristics. We will evaluate how the selection of the liquid metal material (PbLi or Li) and the physical parameters ($B_t$, $B_p^0$, $L$, and $\tau$) affect the severity of the response, as measured by characteristics such as the maximum RMS values of velocity, the induced magnetic field, and induced eddy current:
\begin{eqnarray}
\label{eq:rms1} u_{rms,max}& = & \max_{t>0}(2\textrm{KE}/\rho)^{1/2}, \\ 
\label{eq:rms2} b_{rms,max} &= & \max_{t>0}(2\mu_0\textrm{ME})^{1/2}, \\
\label{eq:rms3} j_{rms,max}&=&\max_{t>0}\left(L^{-1}\int_{-L/2}^{L/2} j^2 dz\right)^{1/2},
\end{eqnarray}
and the overall maximum values of the solution variables within the channel:
\begin{equation}
\label{eq:max} f_{max}\equiv \max_{t>0, -L/2<z<L/2} |f(z,t)|,
\end{equation}
where $f(z,t)$ stands for $u$, $b$, $j$, or $p-p_{z=-L/2}$. The values for the Lorentz force $f_{x,rms,max}$ and $f_{x,max}$ are obtained by multiplying the respective values for $j$ by $B_t$.

The linearity of the equations and boundary conditions \eqref{eq:1da_inc}, \eqref{eq:1db_inc}, \eqref{eq:bc_u_inc}, \eqref{eq:bc_b_inc} means that the solution $u$, $b$ is proportional to the magnitude of the forcing term $dB_p/dt$. The maximum values in \eqref{eq:rms1}-\eqref{eq:rms3} and \eqref{eq:max} must be proportional to $B_p^0$, except for the pressure variation, which must be $\sim \left(B_p^0\right)^2$ (see \eqref{eq:pressure_inc}). The effects of $B_t$, $L$, and $\tau$ are less evident and are investigated numerically.

The problem is solved at $B_p^0=1$~T, with $L \in [0.05, 0.1, 0.15, 0.2, 0.25, 0.3]$~m, $B_t \in [4, 5, 6, 7, 8, 9, 10]$~T, and $\tau \in [0.005, 0.01, 0.015, 0.02, 0.025]$~s. The number of  data points obtained for each fluid is 210. The collected data are used to find, for each characteristic in \eqref{eq:rms1}-\eqref{eq:rms3} and \eqref{eq:max}, the best fit by a single power-law combination:
\begin{equation}
\label{eq:power_fit} f \approx K L^\alpha B_t^\beta \tau^\gamma.
\end{equation}
Log-log multilinear regression is used. 

\begin{table}
\centering
\label{tab:transposed_data}
\begin{tabular}{l|c|c|c|c|c|c|c}

Coefficients& $u_{\mathrm{rms},\max}$ & $u_{\max}$ & $b_{\mathrm{rms},\max}$ & $b_{\max}$ & $j_{\mathrm{rms},\max}$ & $j_{\max}$ & $p_{\max}$ \\
\hline
$\alpha$ ($L$)& $1.148$ & $1.112$ & $1.224$ & $1.255$ & $0.2268$ & $1.043$ & $1.204$ \\
$\beta$ ($B_t$)& $-0.6587$ & $-0.6722$ & $-0.6770$ & $-0.6450$ & $-0.6722$ & $0.2575$ & $-0.5903$ \\
$\gamma$ ($\tau$)& $-0.8709$ & $-0.8599$ & $-0.9357$ & $-0.9301$ & $-0.9346$ & $-0.8654$ & $-0.8651$ \\
K& $0.4682$ & $0.7576$ & $0.02385 $ & $0.03491$ & $5.990 \times 10^{4}$ & $6.025 \times 10^{5}$ & $2.820 \times 10^{4}$ \\
$R^2$ & $0.9964$ & $0.9961$ & $0.9969$ & $0.9972$ & $0.9928$ & $0.9962$ & $0.9950$ \\  \hline

\end{tabular}
\caption{The coefficients of the power-law fitting \eqref{eq:power_fit} for the response severity characteristics computed for PbLi. The calculated coefficients of statistical determination, $R^2$, are also listed. The standard SI units, as elsewhere in the paper, are applied.}
\label{tab:Fit_PbLi}
\end{table}

\begin{table}
\centering
\begin{tabular}{l|c|c|c|c|c|c|c}

Coefficients & $u_{\mathrm{rms},\max}$ & $u_{\max}$ & $b_{\mathrm{rms},\max}$ & $b_{\max}$ & $j_{\mathrm{rms},\max}$ & $j_{\max}$ & $p_{\max}$ \\
\hline
$\alpha$ ($L$) & $0.9958$ & $1.012$ & $1.023$ & $1.044$ & $0.0753$ & $0.9760$ & $1.027$  \\
$\beta$ ($B_t$) & $-0.9507$ & $-0.9314$ & $-0.9773$ & $-0.9391$ & $-0.8955$ & $0.0262$ & $-0.9166$  \\
$\gamma$ ($\tau$) & $-0.9670$ & $-0.9647$ & $-1.005$ & $-0.9906$ & $-0.9851$ & $-0.9647$ & $-0.9642$ \\
K & $0.5718$ & $0.9487$ & $0.006978$ & $0.01150$ & $2.000 \times 10^{4}$ & $3.489 \times 10^{6}$ & $9.324 \times 10^{3}$ \\
$R^2$ & $0.9994$ & $0.9988$ & $0.9998$ & $0.9999$ & $0.9998$ & $0.9981$ & $0.9997$ \\  \hline

\end{tabular}
\caption{The coefficients of the power-law fitting \eqref{eq:power_fit} for the response severity characteristics computed for Li. The coefficients of statistical determination, $R^2$, are also listed. The standard SI units, as elsewhere in the paper, are applied.}
\label{tab:Fit_Li}
\end{table}

\begin{figure}
    \centering
    \includegraphics[width=0.8\textwidth]{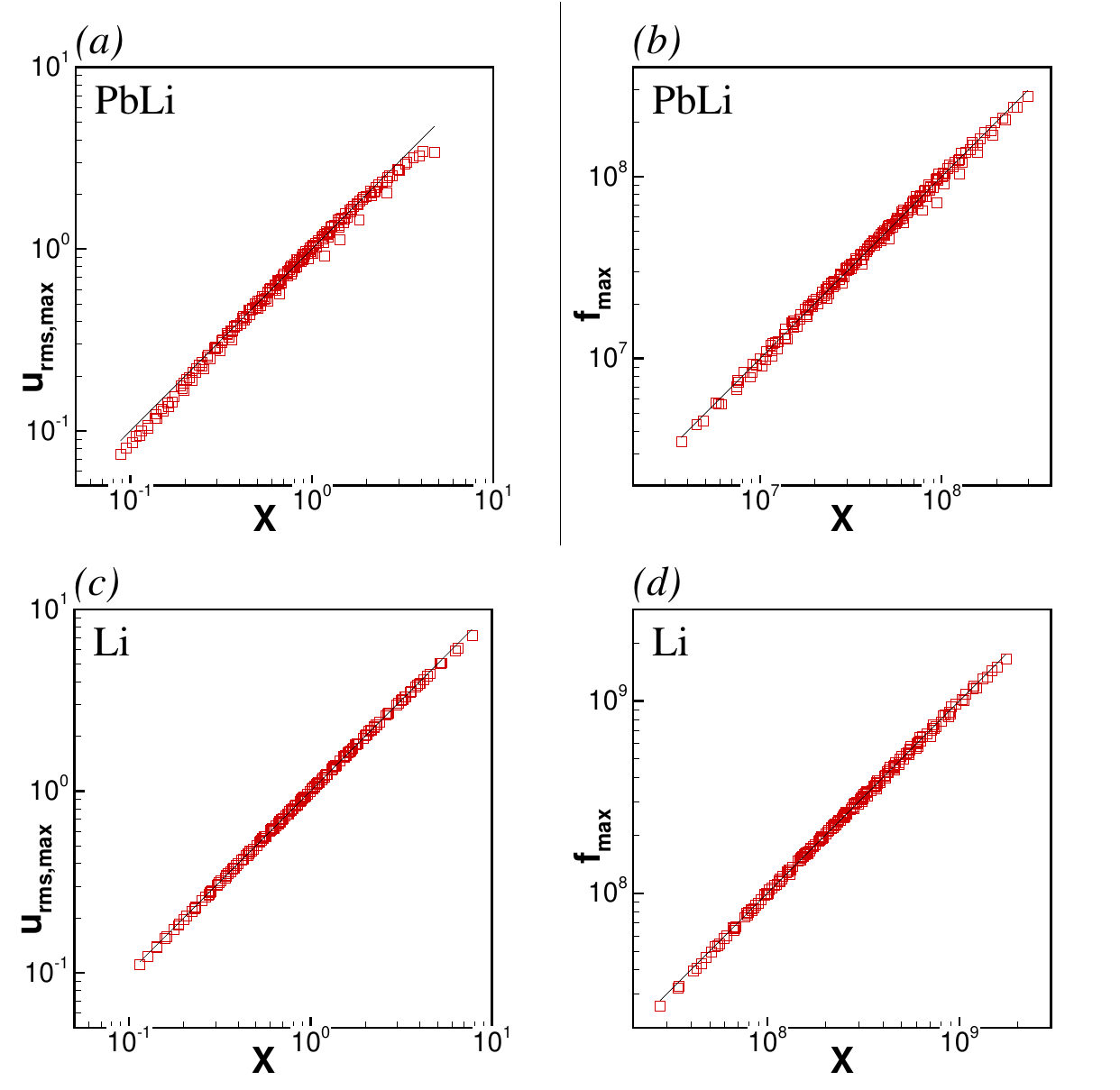}
    \caption{Power-law fit \eqref{eq:power_fit} for $u_{\text{rms,max}}$ and $f_{\text{max}}$ in PbLi and Li flows.  $X=K L^\alpha B_t^\beta \tau^\gamma$ with the coefficients $K$, $\alpha$, $\beta$, and $\gamma$ taken from the respective rows in Tables \ref{tab:Fit_PbLi} and \ref{tab:Fit_Li}.  Symbols represent calculated data, and the straight line denotes the fit.}
    \label{fig:fit}
\end{figure}

The results listed in Tables \ref{tab:Fit_PbLi} and \ref{tab:Fit_Li} and illustrated in Fig.~\ref{fig:fit} show high accuracy of the power-law fits. The value of the coefficient of determination, $R^2$, is at or above 0.993 for all measured characteristics in the PbLi flow case. The accuracy is even higher for the Li flow, with $R^2$ consistently being at or above 0.998.

The data in Tables~\ref{tab:Fit_PbLi} and \ref{tab:Fit_Li} suggest an asymptotic limit, at high $Re$ and $Rm$, for the scaling of the response severity characteristics related to velocity, the induced magnetic field, and pressure:
\begin{equation}
\label{eq:asymptotic} u_{rms,max}\sim b_{rms,max}\sim u_{max}\sim b_{max}\sim \frac{B_p^0 L}{B_t \tau}, \quad p_{max}\sim \frac{\left(B_p^0\right)^2 L}{B_t \tau}.
\end{equation}
No analogous asymptotic limit for the induced current and Lorentz force is indicated by the data.

To explore the possibility of asymptotic scaling, we conducted an additional parametric study for Li with the electrical conductivity $\sigma$ increased by a factor of 20. This increased the magnetic Reynolds number by the same factor, while the hydrodynamic Reynolds number, which was already high in our analysis, remained unchanged. By computing solutions for the same values of $L$, $B_t$, and $\tau$ as before and calculating the power-law fits, we found good agreement (within 0.02 for all coefficients $\alpha, \beta$, and $\gamma$) with the asymptotic scaling \eqref{eq:asymptotic}.

The asymptotic scaling \eqref{eq:asymptotic} is intuitively clear. We have already discussed the effect of the amplitude of the poloidal field $B_p^0$ on the response. The fact that the severity of the response is proportional to $\tau^{-1}$ stems from the appearance of this coefficient in the forcing term of \eqref{eq:1db_inc}. The direct proportionality to the channel width $L$ is consistent with the fact that the total flux of the decaying magnetic field scales as $\sim L$ in our geometry. The presence of $B_t$ in the denominator is consistent with the physical mechanism curtailing the growth of kinetic energy described in Section \ref{sec:typical_PbLi}. The instant of the strongest response occurs when the velocity-induced electric currents $\sim u_{max}B_t$ compensate for the currents induced by the decaying poloidal magnetic field.

A direct explanation based on the governing equations \eqref{eq:1da_inc}, \eqref{eq:1db_inc} can be given for the asymptotic scaling of $u_{rms,max}$ and $u_{max}$. It is based on the observation made in section \ref{sec:typical_solutions} that the threshold states separating the stages of growth and decay of kinetic energy, including the maximum velocity state, are characterized by $b\approx \partial b/\partial t\approx \partial u/\partial t\approx 0$. Applying this approximation and taking the non-diffusive high-$Rm$ limit, we obtain from \eqref{eq:1db_inc}:
\begin{equation}
    \label{eq:reduced_b} \frac{\partial u}{\partial z}\approx \frac{1}{B_t}\frac{dB_p}{dt}=-\frac{B_p^0}{B_t\tau}\exp\left(-\frac{t}{\tau}\right).
\end{equation}
Integrating this equation yields the antisymmetric distribution 
\begin{equation}
\label{eq:u_approx}    u \approx -\frac{B_p^0}{B_t\tau}\exp\left(-\frac{t}{\tau}\right) z,
\end{equation}
which approximates the velocity profiles at the threshold instances, for example, at  $t = 1.055 \times 10^{-3}$ s in the representative PbLi case and at $t=0.251\times 10^{-3}$~s in the representative Li case (see Figs.~\ref{fig:rep_PbLi_profiles}a and \ref{fig:rep_Li_profiles}a). The maximum and RMS magnitudes of the velocity computed from \eqref{eq:u_approx} are:
\begin{equation}
    \label{eq:u_approx_max} |u|_{max} \approx\frac{B_p^0L}{2B_t\tau}\exp\left(-\frac{t_{max}}{\tau}\right), \:\: u_{rms,max} \approx \frac{B_p^0L}{2\sqrt{3}B_t\tau}\exp\left(-\frac{t_{max}}{\tau}\right)
\end{equation}
A comparison of these estimates with the values from the computed solutions for PbLi and Li does not yield good quantitative agreement. At the same time, \eqref{eq:u_approx_max} is fully consistent with the asymptotic scaling \eqref{eq:asymptotic}. 

The parametric study also identifies the system parameters leading to the strongest response. These results are presented in Table \ref{tab:maximums}. For each characteristic of the response severity \eqref{eq:rms1}-\eqref{eq:rms3}, \eqref{eq:max}, we identify the set of parameters ($L$, $B_t$, and $\tau$) for which the maximum is expected and find the respective value in the simulations. A reasonable upper limit of $B_p^0 = 2$~T is taken in all cases.

\begin{table}
    \centering
    \begin{tabular}{l|l|c|c}
      Property   &  $L$~[m/s], $B_t$~[T], $\tau$~[s] & Value for PbLi & Value for Li\\\hline
         $u_{rms,max}$ [m/s]& 
     0.3, 4, $5\times 10^{-3}$& 6.84&14.3\\
 $u_{max}$ [m/s]& 0.3, 4, $5\times 10^{-3}$& 10.8&23.5\\
 $b_{rms,max}$ [T]& 0.3, 4, $5\times 10^{-3}$& 0.517&0.223\\
 $b_{max}$ [T]& 0.3, 4, $5\times 10^{-3}$& 0.748&0.346\\
 $j_{rms,max}$~[A/m$^2$] & 0.3, 10, $5\times 10^{-3}$ & $1.32\times 10^6$ & $4.32\times 10^5$ \\ 
 $j_{max}$~[A/m$^2$]& 0.3, 10, $5\times 10^{-3}$& $5.55\times 10^7$&$3.30\times 10^8$\\
 $p_{max}$~[Pa]& 0.3, 4, $5\times 10^{-3}$& $8.98\times 10^5$&$4.61\times 10^5$\\
  $f_{x,rms,max}$~[N/m$^3$] & 0.3, 10, $5\times 10^{-3}$ & $1.32\times 10^7$ & $4.32\times 10^6$ \\ 
$f_{x,max}$~[N/m$^3$]& 0.3, 10, $5\times 10^{-3}$& $5.55\times 10^8$&$3.30\times 10^9$\\ \hline\end{tabular}
    \caption{Maximum values of the response characteristics determined in the parametric study. For each characteristic, the values of $L$, $B_t$, and $\tau$ that yield the maximum response are listed. All data correspond to $B_p^0 = 2$~T.}
    \label{tab:maximums}
\end{table}

\section{Effect of compressibility}
 {In this section we consider the hypothesis proposed on the basis of dimensional analysis in \cite{smolyanov2025exploratory} that the propagation of pressure waves with a typical time scale comparable to that of the transient plasma event may have a non-negligible effect on the liquid-metal response. The hypothesis is tested using the one-dimensional version of the compressible flow model \eqref{eq:induction-sep}--\eqref{eq:ampere}, \eqref{eq:cont}, \eqref{eq:NStcomp}. Since the propagation of wall-normal pressure waves is now allowed,} the wall-normal velocity $w(z,t)\bm{e}_z$ must be included in the solution. The wall-normal component of $\bm{b}$ must remain zero to satisfy the solenoidality condition \eqref{eq:divb-sep}. The simplified equations for a one-dimensional flow are:

\begin{eqnarray}
\label{eq:1da_comp}\frac{\partial u}{\partial t} +w\frac{\partial u}{\partial z} & = & \nu \frac{\partial^2u}{\partial z^2}+\frac{B_t}{\mu_0 \rho}\frac{\partial b}{\partial z},\\
\label{eq:1db_comp}\frac{\partial w}{\partial t} + w \frac{\partial w}{\partial z} & = & -\frac{1}{\rho}\frac{\partial }{\partial z}\left[p+\frac{b^2}{2\mu_0} \right] + \nu \frac{\partial^2 w}{\partial z^2} - \frac{B_p(t)}{\mu_0\rho}\frac{\partial b}{\partial z},\\
\label{eq:1dc_comp}\frac{\partial b}{\partial t} + w\frac{\partial b}{\partial z}  & = & B_t \frac{\partial u}{\partial z}+\eta \frac{\partial^2 b}{\partial z^2}+\frac{B_p^0}{\tau}\exp\left(-\frac{t}{\tau}\right),\\
\label{eq:1dd_comp}\frac{\partial p}{\partial t} + w\frac{\partial p}{\partial z}  & = & -\rho C^2 \frac{\partial w}{\partial z}.
\end{eqnarray}
The gradients of the thermodynamic and magnetic pressures on the right-hand side of \eqref{eq:1db_comp} are combined to represent the gradient of the cumulative pressure, $\Pi=p+b^2/(2\mu_0)$.

The boundary conditions \eqref{eq:bc_b_inc} on the magnetic field remain valid. The selection of the hydrodynamic boundary conditions is based on the typical values of the acoustic impedance, $Z\equiv \rho C$, of the liquid and the wall. Typical wall materials for the poloidal duct in reactor blankets, namely steel and SiC ceramic, have an impedance of $\sim 4\times 10^7\text{ kg}/(\text{m}^2\text{s})$, which is roughly three times larger than that of liquid PbLi and about twenty times larger than that of liquid Li. This allows us to use, as an approximation suitable for this initial study of the effect, the model of a stiff wall. The resulting boundary conditions are:
\begin{equation}
    \label{eq:bc_p_comp} \frac{\partial p}{\partial z} = u=w=b=0, \:\: \textrm{at} \:\: z=\pm \frac{L}{2}.
\end{equation}

The problem is solved numerically using the same method as for the incompressible flow problem, but with minor adjustments introduced to avoid spurious oscillations and numerical instability caused by fast-propagating pressure waves. The spatial grid resolution is increased to 2000 points, while the clustering constant (see \eqref{eq:tanh}) is reduced to $A=3.0$. Furthermore, to allow the direct application of the pressure boundary conditions \eqref{eq:bc_p_comp} and ensure numerical stability, an artificial diffusivity term, $\kappa \partial^2 p/\partial z^2$, is added to the right-hand side of \eqref{eq:1dd_comp}. The small value $\kappa=10^{-4}\text{ m}^2/\text{s}$ ensures that the effect of this term on the solution is negligible. This is confirmed by additional simulations showing practically no difference when $\kappa$ is increased to $10^{-3}\text{ m}^2/\text{s}$.

\begin{figure}
    \centering
    \includegraphics[width=0.8\textwidth]{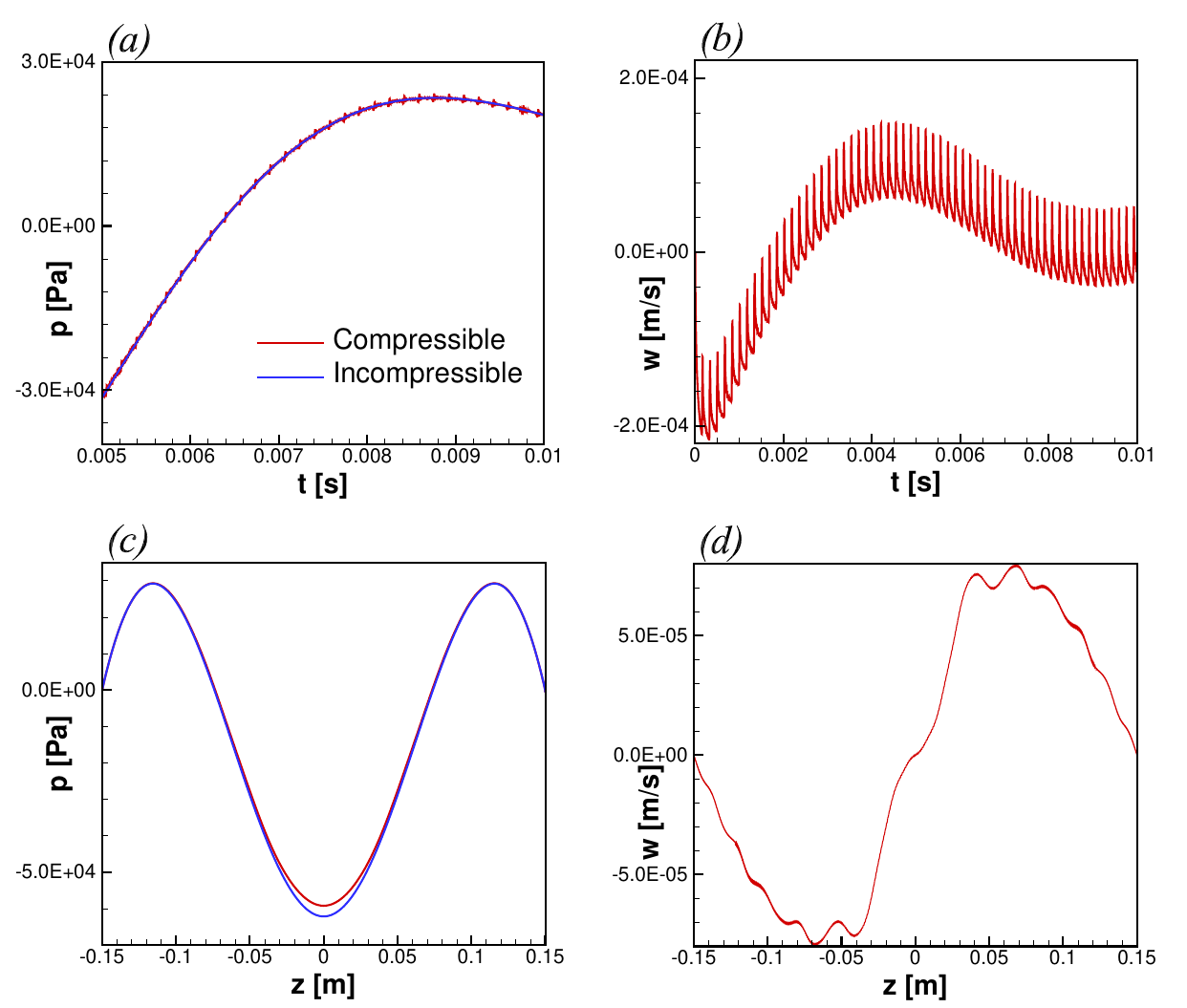}
    \caption{Effect of compressibility on the flow evolution. Solutions of the compressible \eqref{eq:1da_comp}–\eqref{eq:1dd_comp} and incompressible \eqref{eq:1da_inc}–\eqref{eq:1db_inc} problems obtained for PbLi at $L=0.3\text{ m}$, $B_t=4\text{ T}$, $B_p^0=2\text{ T}$, and $\tau=0.005\text{ s}$ are presented. Time signals of the pressure $p$ and wall-normal velocity $w$ at $z=-0.45L$ are shown in \emph{(a)} and \emph{(b)}, respectively. Only a portion of the computed evolution is included in \emph{(a)} for better visibility. Snapshots of profiles of $p$ and $w$ at $t=0.01\text{ s}$ are shown in \emph{(c)} and \emph{(d)}. The wall-normal velocity is present only in the compressible flow case. }
    \label{fig:comp}
\end{figure}

The solutions were computed for PbLi and Li flows using the parameter set for which the strongest response was identified in the incompressible flow case (see Table \ref{tab:maximums}): $L=0.3\text{ m}$, $B_t=4\text{ T}$, $B_p^0=2\text{ T}$, and $\tau=0.005\text{ s}$. The results obtained for PbLi are illustrated in Fig.~\ref{fig:comp}, with the incompressible flow solution included for comparison.

Non-zero wall-normal velocities $w$ and pressure perturbations are observed. Their identification as pressure waves is confirmed by the perfect agreement between the oscillation period, $T_{\text{osc}}=0.168\times 10^{-3}\text{ s}$, of the fluctuations (computed as the interval between subsequent peaks of $p$ and $w$ in the time signals in Fig.~\ref{fig:comp}a,b) and the travel time $L/C$ of the pressure waves across the domain. At the same time, the amplitude of these perturbations remains small. Analysis of other flow characteristics (longitudinal velocity $u$, induced magnetic field $b$, kinetic and magnetic energies, etc.) reveals no significant effect of compressibility.

Similar results are obtained for the flow of Li, except that the amplitude of the pressure waves is even weaker in that case.  {We conclude that for the system analyzed in this study, the hypothesis proposed in \cite{smolyanov2025exploratory} does not hold.} Pressure waves develop and can be detected, but their effect on the overall flow evolution is negligibly small.

\section{Conclusion}\label{sec:conclusion}
 {We have analyzed the response of a liquid metal confined in a channel between two non-conducting walls to a rapid variation of the imposed magnetic field. While simplified, our one-dimensional model captures the key physical effects of the process. The choice of configuration, system parameters, and physical properties of the fluid (PbLi or Li melts) makes the analysis relevant to key processes expected in a poloidal duct of a fusion reactor blanket during a transient plasma event.}

 {While this study is directly relevant to the specific configuration of a poloidally oriented channel and the typical time variation of the magnetic field during a plasma disruption, several factors of a real blanket system are not captured by our model. These include, for example, the actual three-dimensional geometry of the blanket module, which includes multiple ducts and inlet/exit manifolds, a finite wall conductivity, the three-dimensional structure of the magnetic field, and its actual time history, which may differ from the exponential decay assumed in our work. However, the underlying physical mechanisms revealed by our analysis are general and are expected to play a major role in response of an actual blanket to a plasma disruption.}

 {The main result of our work is a clear and comprehensive explanation of the underlying physics of the process. Most importantly, our analysis confirms the hypothesis proposed on the basis of a reduced model in \cite{smolyanov2025exploratory} that the response of a liquid metal to a plasma disruption event has an inherent oscillatory component. The oscillations are identified as a manifestation of forced standing Alfvén waves that develop as the liquid metal moves across the lines of the toroidal magnetic field. This simple physical explanation, proposed here for the first time, is confirmed by the close agreement between the computed oscillation periods and those theoretically predicted for Alfvén waves.}

 {The realization of the Alfvén wave mechanism in an actual fusion reactor blanket is currently unknown and would be an interesting direction for future work. The fundamental nature of the mechanism allows us to predict existence of the waves in an actual blanket, but their dynamics are three-dimensional and complex, and are likely to involve interaction of multiple Alfvén modes. It is also possible that the oscillatory behavior in realistic blanket geometries may be less pronounced than in the idealized channel considered here.}

 {A practical contribution of our work to the design of fusion reactor blankets is the quantitative characterization of the response. The data indicate that the response is severe and rapid. On a millisecond timescale, the fluid is accelerated to velocities up to $10\text{--}20\text{ m/s}$, and eddy currents as high as $10^8\text{ A/m}^2$ are generated. The system may experience Lorentz forces of $\sim 10^9\text{ N/m}^3$, an induced magnetic field of $\sim 0.5\text{ T}$, and pressure variations of nearly $10^6\text{ Pa}$. Obtained for an idealized system, these values must not be taken as a direct forecast of the behavior of an actual blanket. However, even when treating them as upper-bound estimates and assuming values an order of magnitude lower for a real system, we conclude that the consequences for processes within the blanket and the reactor as a whole are likely to be significant.}

 {We have also explored the possibility, suggested on the basis of dimensional analysis in \cite{smolyanov2025exploratory}, that high-amplitude pressure variations and the extremely short timescale of the process may result in a significant influence of pressure waves on the flow. Simulations conducted within the framework of a linearized barotropic compressibility model show that, for blanket-relevant parameters, the pressure waves are detectable, but their effect on the flow evolution is negligibly small.}

Several directions for future work appear promising. One of them aligns with the central goal of the SciDAC-5 program supporting our study: the development of comprehensive numerical simulation tools that include the transient plasma and the liquid-metal blanket as fully coupled parts of a single system. The results of our study provide a useful benchmark for such models. Another direction involves computational solution of model problems to explore how the physics of the process is affected by various features of the blanket system absent in our current model. Yet another important direction belongs to the realm of reactor design rather than fluid mechanics. The results of our analysis can be used to predict, and safeguard against, the negative effects of transient plasma events on the operation and structural integrity of the reactor.

\begin{acknowledgments}
This work was performed under the SciDAC-5 project “Center for Simulation of Plasma–Liquid Metal Interactions in Plasma Facing Components and Breeding Blankets of a Fusion Power Reactor,” supported by the U.S. Department of Energy. Oak Ridge National Laboratory acknowledges support from the U.S. Department of Energy under contract DE-AC05-00OR22725. The University of Michigan–Dearborn acknowledges support through a subcontract with Oak Ridge National Laboratory under contract CW52278.

This manuscript has been authored by UT-Battelle, LLC, under contract DE-AC05-00OR22725 with the U.S. Department of Energy. The U.S. Government retains, and the publisher, by accepting the article for publication, acknowledges that the U.S. Government retains, a nonexclusive, paid-up, irrevocable, worldwide license to publish or reproduce the published form of this manuscript, or to allow others to do so, for U.S. Government purposes. The U.S. Department of Energy will provide public access to these results of federally sponsored research in accordance with the DOE Public Access Plan (https://www.energy.gov/doe-public-access-plan).
\end{acknowledgments}

\providecommand{\noopsort}[1]{}\providecommand{\singleletter}[1]{#1}%

\end{document}